# A Predictive Model of Geosynchronous Magnetopause Crossings

A. Dmitriev, A. Suvorova, J.-K. Chao

```
FORTRAN code of predictive model of geosynchronous magnetopause crossings (GMC)
Dmitriev, A., A. Suvorova, and J.-K. Chao (2011), A predictive model of
geosynchronous magnetopause crossings, J. Geophys. Res., 116, A05208,
doi:10.1029/2010JA016208.
Determining solar wind pressure Pdm required for GMC
at given location in aberrated GSM system:
LT & Lat
and under given:
Bz & Dst

Written by Alexei Dmitriev <alexei_dmitriev@yahoo.com>

      subroutine GMCmod
     *(
     *  vLT,    ! aGSM local time (h)
     *  vLat,   ! aGSM latitude (deg)
     *  Dst,    ! 1-min Dst variation (nT)
     *  Bz,     ! aGSM Bz (nT)
     *  Pdm     ! modelled total SW boundary pressure (nPa)
     *)

      pi=acos(-1.)
      s=sin((vLT-12.)*15*pi/180.)
      s2=s*s
      s3=s2*s
      s4=s3*s
      sLat=sin(abs(vLat)*pi/180.)
      sLat2=sLat*sLat

vLT->{b, Bz0, Pmax, Pmin}
      b=0.142621-0.018836*s
      Bz0=1.88986+3.61671*s-8.22981*s2

Pmax&Pmin=F{vLT}
      Pmax=17.3+70.2*s2
      Pmin=5.56673+1.73167*s+7.60775*s3+15.7509*s4

Lat
      if(vLT.ge.9.and.vLT.le.15) then
        Pmax=Pmax+(10**1.69873)*(sLat**1.65733)
        Pmin=Pmin+(10**1.31338)*(sLat**1.92242)
      endif

Dst
      Pmax=Pmax+6.99418*exp(0.0094883*Dst)
      Pmin=Pmin+8.68686*exp(0.0174171*Dst)

      bDst=0.2
      if (Dst.ge.-150) bDst=0.0739-Dst/1300.
      b=sqrt(abs(b)*abs(bDst))

      Pdm=Pmax-(Pmax-Pmin)/(1.+exp(b*(Bz+Bz0)))

      return
      end
```

# A Predictive Model of Geosynchronous Magnetopause Crossings

A. Dmitriev[1,2], A. Suvorova[2], J.-K. Chao[1]

[1]*Institute of Space Science National Central University, Chung-Li, Taiwan*

[2]*Institute of Nuclear Physics Moscow State University, Moscow, Russia*

———

A. V. Dmitriev, Institute of Space Science National Central University, Chung-Li, 320, Taiwan, also D.V. Skobeltsyn Institute of Nuclear Physics, Moscow State University, Russia (e-mail: dalex@jupiter.ss.ncu.edu.tw)

A. V. Suvorova, D.V. Skobeltsyn Institute of Nuclear Physics, Moscow State University, Russia (e-mail: suvorova_alla@yahoo.com)

J.-K. Chao, Institute of Space Science National Central University, Chung-Li, 320, Taiwan (e-mail: jkchao@jupiter.ss.ncu.edu.tw)

**Abstract** We have developed a model predicting whether or not the magnetopause crosses geosynchronous orbit at given location for given solar wind pressure *Psw*, *Bz* component of interplanetary magnetic field (IMF) and geomagnetic conditions characterized by 1-min *SYM-H* index. The model is based on more than 300 geosynchronous magnetopause crossings (GMCs) and about 6000 minutes when geosynchronous satellites of GOES and LANL series are located in the magnetosheath (so-called MSh intervals) in 1994 to 2001. Minimizing of the *Psw* required for GMCs and MSh intervals at various locations, *Bz* and *SYM-H* allows describing both an effect of magnetopause dawn-dusk asymmetry and saturation of *Bz* influence for very large southward IMF. The asymmetry is strong for large negative *Bz* and almost disappears when *Bz* is positive. We found that the larger amplitude of negative *SYM-H* the lower solar wind pressure is required for GMCs. We attribute this effect to a depletion of the dayside magnetic field by a storm-time intensification of the cross-tail current. It is also found that the magnitude of threshold for *Bz* saturation increases with *SYM-H* index such that for small negative and positive *SYM-H* the effect of saturation diminishes. This supports an idea that enhanced thermal pressure of the magnetospheric plasma and ring current particles during magnetic storms results in the saturation of magnetic effect of the IMF *Bz* at the dayside magnetopause. A noticeable advantage of the model prediction capabilities in comparison with other magnetopause models makes the model useful for space weather predictions.

# 1. Introduction

Magnetopause crossings of geosynchronous orbit located at distance of ~6.6 Earth's radii (Re) occur very rare and require very strong disturbances in the solar wind (SW) and magnetosphere [*Russell*, 1976; *Rufenach et al.*, 1989; *McComas et al.*, 1993; *Suvorova et al.*, 2005; *Li et al.*, 2010]. In spite of rarity, they result in prominent and dramatic space weather effects such as a damage or even loss of geosynchronous satellites [e.g. *Odenwald and Green*, 2007]. Only few magnetopause models adopt geosynchronous magnetopause crossings (GMCs) and, thus, enable predicting the magnetopause under strongly disturbed conditions [*Kuznetsov and Suvorova*, 1998a; *Shue et al.*, 1998; *Dmitriev and Suvorova*, 2000; *Yang et al.*, 2003; *Lin's et al.*, 2010]. Prediction of GMCs by those models gives sometimes controversial results [e.g. *Shue et al.*, 2000; *Ober et al.*, 2002; *Yang et al.*, 2002; *Suvorova et al.*, 2005; *Dmitriev et al.*, 2005].

In modeling the magnetopause at geosynchronous orbit, the following problems rise up [e.g. *Shue et al.*, 2000; *Yang et al.*, 2003; *Suvorova et al.*, 2005]: orbital bias, GMC identification, dependence on upstream conditions. For strongly disturbed SW conditions, the magnetopause location varies significantly but geosynchronous satellites are located at the fixed radial distance. That leads to a problem of orbital bias. Namely, when a geosynchronous satellite is situated in the magnetosheath (MSh), the magnetopause distance can be much smaller than 6.6 Re. Hence, if a model is based on MSh observations, the upstream SW conditions for GMCs can be overestimated, i.e. such a model necessarily predicts stronger SW pressure or more negative IMF *Bz* than those are actually required. And vice versa, a model based on magnetosphere observations, underestimates the SW conditions required for GMCs.

Identification of magnetopause crossings at geosynchronous orbit is performed using magnetic measurements onboard GOES satellites [e.g. *Rufenach et al.*, 1989] and plasma measurements onboard LANL satellites [e.g. *McComas et al.*, 1993]. It is rather difficult to identify the magnetopause from magnetic measurements during large northward IMF, when magnetic field in the magnetosheath almost coincides with the magnetospheric field. That results in a high rate of false alarms for the sets of GMCs collected from GOES data. For instance, under very high SW pressure and northward IMF, a model can correctly predict the inbound magnetopause crossing, but that crossing is not revealed in GOES magnetic data. In this case the correct model prediction is wrongly assigned to the false alarm. GMC identification using LANL plasma data becomes difficult during strong enhancements of solar energetic particles, which often accompany geomagnetic storms [*Dmitriev et al.*, 2005]. The radiation effect causes missing the magnetopause crossings and, thus, increases the false alarms by mistake. Accurate identification of the GMCs requires developing sophisticated methods for magnetic and plasma data analysis [e.g. *Suvorova et al.*, 2005].

In contrast to magnetopause crossings under quiet and slightly disturbed conditions, the GMCs are characterized by such non-linear effects as IMF $Bz$ influence saturation, dawn-dusk asymmetry and, perhaps, preconditioning by the IMF $Bz$. The effect of saturation was studied by a few authors [*Kuznetsov and Suvorova*, 1994; 1996; *Shue et al.*, 1998; *Dmitriev and Suvorova*, 2000; *Dmitriev et al.*, 2001b; *Yang et al.*, 2003; *Suvorova et al.*, 2005]. It was found that for a given SW pressure and strong southward IMF, the magnetopause distance stops decreasing when the magnitude of negative $Bz$ exceeds a certain threshold. The threshold for saturation was estimated to be about -20 nT and its magnitude increases with the SW pressure [*Yang et al.*, 2003].

The nature of saturation effect is still under investigation. On the base of global MHD simulations, the decrease and limitation of the reconnection at the dayside magnetopause can result from a magnetic effect of the region 1 field-aligned current [*Siscoe et al.*, 2004] and/or from high-density plasmaspheric plasma flowing into the reconnection site [*Borovsky et al.*, 2008]. Based on statistical analysis of GMCs, *Suvorova et al.* [2003] proposed that the reconnection saturation might be caused by an enhanced thermal pressure of the magnetospheric plasma and ring current particles during strong magnetic storms.

Dawn-dusk asymmetry of GMCs was reported and studied in a number of papers [*Wrenn et al.*, 1981; *Rufenach et al.*, 1989; *McComas et al.*, 1993; *Itoh and Araki*, 1996; *Kuznetsov and Suvorova*, 1996; 1997; 1998b; *Dmitriev and Suvorova*, 2000; *Dmitriev et al.*, 2004; 2005]. It was found that the GMCs occur mostly often in prenoon sector and, thus, the magnetopause is closer to the Earth on the dawnside than on the duskside. This skewing can be represented by a rotation of the magnetopause nose point toward dawn by an angle of ~15° [*Dmitriev et al.*, 2004] or rather by a shifting of the magnetopause toward dusk by few tenths to 2 Re [*Kuznetsov and Suvorova*, 1998a,b; *Dmitriev et al.*, 2005]. The dawn-dusk asymmetry, increasing with southward IMF and with geomagnetic storm activity, is attributed to storm-time intensification of the asymmetrical ring current peaking in the dusk sector. However there are no any numerical descriptions of that asymmetry.

An importance of the effect of preconditioning by the IMF *Bz* was pointed out by *Shue et al.* [2000]. It is suggested that longer duration of southward IMF promotes closer approach of the magnetopause to the Earth. This requires consideration of an integral effect of the IMF *Bz*. Modeling this effect is quite difficult because it is non-linear in time, i.e. during integration, consequent magnitudes of *Bz* might have different weights depending on time.

As we can see from the above, modeling the GMCs is very complex procedure, which has to take into account several spatial and temporal nonlinear effects. We have to point out that those specific effects become vanishingly small under quiet or moderately disturbed conditions. Because of that, all extrapolations of the magnetopause from the moderately disturbed conditions to the range of GMCs were failed. In addition, the determination of magnetopause crossings is still not very accurate especially for northward IMF. Even after firm determination of a GMC, we can only say about relative location of the magnetopause: inside or outside of geosynchronous orbit; the distance to magnetopause can not be determined precisely. Hence, a new model approach for such events is required.

In the present study we develop a method for prediction the GMCs. Instead prediction of the magnetopause distance, we build a model, which says weather or not the magnetopause crosses geosynchronous orbit at given location and for given solar wind and magnetospheric conditions. Experimental data used for modeling are presented in Section 2. In Section 3 we introduce the method and construct the model. Section 4 is devoted to the model evaluation and comparison with other models. The results are discussed in Section 5. Section 6 is conclusions.

## 2. Experimental data

Geosynchronous magnetopause crossings are selected in the time period from 1994 to 2001 by using magnetic field and plasma data acquired from geosynchronous satellites GOES 8, 9, 10 and LANL 1990-095, 1991-080, 1994-084, LANL-97A, 1989-046, respectively. The satellites occupy all the longitudinal sectors as shown in Figure 1. That allows smoothing so–called longitudinal and latitudinal effects [*Suvorova et al.*, 2005]. The former one consists in ~10% variation of the geodipole field magnitude with geographic longitude in the equatorial plane. The latitudinal effect is owing to higher probability for a geosynchronous satellite to be located at middle GSM latitudes of ~20°. In Figure 2 one can see that the GMCs are scattered quite uniformly around the subsolar point. We can only indicate a dawn-dusk asymmetry of the distribution such that the number crossings is higher in the dawn and prenoon sectors than that in postnoon and especially dusk sectors. Note that this asymmetry is a natural phenomenon characterizing the strongly disturbed dayside magnetosphere [e.g. *Dmitriev et al.*, 2004]. We will discuss this effect later.

The method of GMC selection is described in detail by *Suvorova et al.* [2005]. Briefly, we used high-resolution (~1 minute) ISTP data (http://cdaweb.gsfc.nasa.gov/cdaweb/istp_public/) from geosynchronous satellites GOES and LANL and upstream monitors Geotail, Wind, and ACE. We analyzed so-called magnetosheath intervals (hereafter MSh intervals), when a geosynchronous satellite was located in the magnetosheath. A MSh interval was identified using the GOES magnetometer data, when one of two requirements were satisfied: 1) the magnetic field measured by the GOES deviated significantly from the geomagnetic field with dominant northward component in the dayside magnetosphere; 2) the magnetic field $Y$ and $Z$ GSM components measured by the GOES correlated with the corresponding IMF components measured by an upstream monitor. Identification of MSh intervals from the LANL data was based on a substantial increase of the low energy ion and electron content proper to the magnetosheath conditions. The satellite location is ordered in a fully aberrated GSM (aGSM) coordinate system where the $X$-axis is anti-parallel to the solar wind velocity.

Each MSh interval for each geosynchronous satellite was considered as a case event, for which we determined the corresponding upstream solar wind conditions. The timing was based on a time delay for direct propagation of the entire solar wind structure observed by an upstream monitor, and on an additional time shift, which is owing to tilted interplanetary fronts. The final timing was verified using two independent criteria. The first general criterion was based on a correlation between the SW pressure and the *Dst* (*SYM-H*) index that originates from SW pressure-associated changes of the Chapman-Ferraro current at the magnetopause [e.g., *Burton,* 1975; *Russell et al.*, 1994a,b]. The best correlation indicates the best timing as well as the best choice of the upstream solar wind monitor. The second, subsidiary, criterion for the upstream solar wind timing for the GOES satellites is co-variation of the magnetic field *Y* and *Z* GSM components measured by GOES during the magnetosheath intervals with the corresponding IMF components measured by an upstream solar wind monitor. This method is similar to a method of clock angle [e.g. *Song et al.*, 1992], which is a function of magnetic field *Y* and *Z* components. However, the method of clock angle works well only in subsolar region. Considering *Y* and *Z* components separately enable us to analyze a wide dayside spatial range. For the LANL satellites, the independent criterion is the co-variation of the ion density in the magnetosheath with the SW pressure.

In such a way we select only events, for which the magnetopause dynamics is directly associated with variations of the solar wind conditions or with changing of the geosynchronous satellite location. The accuracy of such methods, based on ~1-minute time resolution of the experimental data, is estimated as a few minutes. The accuracy can be affected by a non-instantaneous magnetopause response, continuous changes of the solar wind front tilt [e.g. *Collier et al.*, 1998], rapid variations of the solar wind plasma and IMF properties, and the evolution of the SW irregularities propagating through the interplanetary medium and the magnetosheath [*Richardson and Paularena*, 2001; *Weimer et al.,* 2002].

As a result, we have identified 129 and 197 magnetosheath entries (or GMCs) for the GOES and LANL satellites, respectively. The MSh intervals observed by the GOES and LANL satellites contain, respectively, 3004 and 2851 measurements. These statistics allowed studying such important features of the geosynchronous magnetopause crossings as dawn-dusk asymmetry and IMF *Bz* influence saturation.

The dawn-dusk asymmetry of GMCs is demonstrated in Figure 3. Owing to magnetopause flaring, higher pressure is required to produce crossings on the flanks than that in noon region. For large positive *Bz* (>5 nT) the scatter plot of *Psw* versus MLT is practically symmetric. The GMCs and MSh intervals associated with the minimal required SW pressure *Psw* ~ 15 nPa are observed around noon, while a very high pressure *Psw* ~ 70 nPa is observed for GMCs at 0700 MLT or at 1700 MLT. For whole statistics, including negative *Bz*, the situation changes dramatically. The scatter plot of *Psw* versus MLT demonstrates very strong dawn-dusk asymmetry: the minimal SW pressures *Psw* ~ 5 nPa are observed mostly in the prenoon sector at ~1000 MLT. At flanks, the minimal SW pressure required for the GMC is about 10 nPa at 0700 MLT and *Psw* ~ 30 nPa at 1700 MLT, i.e. as much as 3 times or more. Such a difference is manifestation of substantial duskward skewing of the dayside magnetopause during strongly disturbed conditions accompanied by large southward IMF.

The IMF *Bz* influence saturation can be revealed in a scatter plot of GMCs in a space of the SW pressure versus IMF *Bz* presented in Figure 4. One can see that the solar wind conditions for GMCs, varying over a very wide range of *Psw* (from ~4 nPa to >100 nPa) and *Bz* (from -40 nT to 40 nT), are restricted rather sharply by a lower envelope boundary, below which GMCs are not observed, excepting a few outliers. The envelope boundary corresponds to minimal solar wind conditions, which are necessary for GMCs. Numerically the boundary can be represented by the following expression [*Suvorova et al.*, 2005]:

$$Psw = 21 - \frac{16.2}{1 + \exp\{0.2(Bz - 2.)\}} \qquad (1)$$

The right horizontal branch of the envelope boundary, asymptotically approaching to *Psw* = 21 nPa, corresponds to a regime of pressure balance for the magnetopause under strong northward IMF. The left branch approaches the *Psw* ~ 4.8 nPa under very strong negative *Bz* and is associated with the regime of *Bz* influence saturation. In that regime, the increasing of southward IMF above the threshold of ~-20 nT does not affect the magnetopause location.

In Figure 4 one can see that different magnetopause models predict very different solar wind conditions required for GMCs. For positive *Bz*, the predicted solar wind pressure varies from ~26 nPa for *Lin's et al.* [2010] model to ~45 nPa for *Yang's et al.* [2003] model. In the regime of saturation for strong negative *Bz* the difference is also big: from 4.6 nPa for *Lin's et al.* [2010] to ~8 nPa for *Shue's et al.* [1998] model. Note that *Petrinec and Russell's* [1996] model is unable predicting the saturation. We have to point out that most of the models require stronger SW pressure for GMCs than that derived for the lower envelope boundary (Equation 1). Hence, prediction of the subsolar magnetopause by the existing models is quite ambiguous.

## 3. Modeling the GMCs

A crucial problem in modeling the GMCs is the effect of orbital bias. In order to minimize this effect, *Kuznetsov and Suvorova* [1998a] proposed a method for selection of the GMCs. The method is based on determination of a surface of minimal conditions required for inbound magnetopause crossings in the three-dimensional space of SW pressure, IMF *Bz*, and local time. For every value of IMF *Bz* and local time, the lowest value of the SW pressure is determined. It was reasonably assumed that the SW conditions selected in such a way are the ones required for the equilibrium magnetopause location just near geosynchronous orbit.

In the present study we extend the method of *Kuznetsov and Suvorova* [1998a]. We will determine the envelope boundary for GMCs and MSh intervals at scatter plot of *P*sw versus *Bz* for various MLT, aGSM latitudes and geomagnetic conditions characterized by *SYM-H* index (1-min equivalent of *Dst* index). In modeling we consider MSh intervals and inbound magnetopause crossings, i.e. when a geosynchronous satellite entries to the magnetosheath. We do not model outbound magnetopause crossings when a geosynchronous satellite returns to the magnetosphere because for them the SW pressure and/or IMF *Bz* can be much smaller than required for GMCs.

The envelope boundary is fitted by a hyperbolic tangent function:

$$Psw(Bz) = P_{max} - \frac{P_{max} - P_{min}}{1 + \exp\{\chi(Bz + Bz_0)\}} \qquad (2)$$

The variables $P_{max}$ and $P_{min}$ are estimated as asymptotes of the function $Psw(Bz)$ when $Bz \to +\infty$ and $Bz \to -\infty$, respectively. The coefficients $\chi$ and $Bz_0$ characterize, respectively, the steepness of inflection and inflection point of the envelope boundary. They can be calculated by a simple approximation to the points located in the close vicinity of the boundary. The parameters of Equation 2 are considered as functions of aGSM longitude (or MLT), latitude and *SYM-H* index. Note that Equation (2) can be linearized relative to *Bz*: $\log(Psw - P_{min}) - \log(P_{max} - Psw) = \chi \cdot (Bz+Bz_0)$. Hence, the solar wind pressure *Psw* should be represented in logarithmic scale.

*3.1. Dependence on MLT*

As a first step, we study how the envelope boundary changes with MLT. Here we should take into account the effect of dawn-dusk asymmetry (see Figure 3). The lowest SW pressures are observed in the range from 10 to 12 MLT. Hence, the ranges of MLT will not be symmetric around noon. Table 1 shows the MLT grid used for modeling. An example of envelope boundary determination in the range from 13 to 18 MLT is presented in Figures 5 and 6. The method of boundary determination is described in detail in *Suvorova et al.* [2005].

Namely, we analyze a two-dimensional (*Psw* versus *Bz*) distribution of occurrence the GMCs and MSh intervals (Figure 6). The space of parameters is split into 20 x 20 bins with width d*Bz* = 3 nT and height increasing logarithmically with *Psw*. For each bin the number of magnetosheath points and GMCs (the latter is weighted by a factor of 3) are summed. We weight the GMCs more heavily because during selection of the MSh intervals the magnetosheath entry is considered as a reference point. To select the meaningful events, we adopt an occurrence number of 5 as a lower threshold for meaningful statistics. The envelope boundary corresponds to the statistically significant bins with the lowest SW pressure for each given *Bz*. Below this boundary the occurrence number decreases sharply from ≥5 (gray shading) to <5 (white). By this way we estimate the asymptotes $P_{max}$ and $P_{min}$ (see Table 1) and select the points for approximation of the envelope boundary.

In Figure 5 one can see that the lower boundary shifts toward higher pressures relative to the lowest boundary, derived for all GMCs. The shift is due to blunted shape of the magnetopause such that higher pressure is required for GMC at larger MLT displacement from noon. However the real situation is more complicated because of magnetopause duskward skewing under southward IMF. The skewing causes larger change for $P_{min}$ than for $P_{max}$ relative to their lowest values. From Table 1 we find that in the range of 13 to 18 MLT the $P_{min}$ = 7 nPa, i.e. that is about 50% larger than the lowest value of $P_{min}$ = 4.8 nPa. The $P_{max}$ increases only slightly from the lowest value of 21 nPa to 22 nPa.

A dependence of the $P_{max}$ and $P_{min}$ on aGSM longitude (*mLon*) is presented in Figure 7. Note that the *mLon* is related to MLT as *mLon* = 15°· (MLT-12). The $P_{max}$ is distributed almost symmetrically around the noon, while the dependence for $P_{min}$ is skewed dawnward. We fit the dependencies for $P_{max}$ and $P_{min}$ by polynomial functions of sin(*mLon*):

$$P_{max}(mLon) = 17.3 + 70.2 \cdot \sin^2(mLon) \qquad (3a)$$

$$P_{min}(mLon) = 5.57 + 1.73 \cdot \sin(mLon) + 7.61 \cdot \sin^3(mLon) + 15.8 \cdot \sin^4(mLon) \qquad (3b)$$

From Table 1 one can see that the envelope boundaries are fitted quite well (with correlation coefficient r > 0.8) in practically all MLT ranges. Note that for symmetrical case of large positive *Bz* (Equation 3a), the shape of subsolar magnetopause is close to sphere and the dependence of pressure should be close to $\sin^2(mLon)$. In the asymmetrical case the magnetopause is represented by an expansion into series of sin(*mLon*).

Using hydro-dynamical approach for pressure balance at the magnetopause, one can estimate that the SW pressure required for GMC in subsolar point is somewhere between *Psw* = 20.9 nPa and 47 nPa.

The lower and upper limits correspond to planar and spherical magnetopause, respectively. From Equation 3 we find that in the subsolar point $P_{min}$ = 5.57 nPa and $P_{max}$ = 17.3 nPa. That is smaller then the pressure balance prediction. Note that the latest MP model by *Lin et al.* [2010] predicts for subsolar GMCs the lower and upper SW pressures of $P_{min}$ ~ 5 nPa and $P_{max}$ ~ 26 nPa, respectively, under large negative and positive *Bz*. These inconsistencies might be owing to effects of geomagnetic field depletion by the magnetospheric currents intensified during magnetic storms. We will discuss this problem later.

Figure 8 shows variations of the parameters $\chi$ and $Bz_0$ with the *mLon*. The dependencies of the steepness $\chi$ and inflection point $Bz_0$ on *mLon* can be expressed as the following:

$$Bz_0(mLon) = 1.89 + 3.62 \cdot \sin(mLon) - 8.23 \cdot \sin^2(mLon) \quad (4)$$

$$\chi(mLon) = 0.143 - 0.0188 \cdot \sin(mLon) \quad (5)$$

We have to point out a wide spreading of the parameter $\chi$ in the dawn and prenoon sectors. The spreading results from relatively poor statistics in these regions as one can see in Figure 3. However, we can indicate a negative trend for the steepness $\chi$. It means that for low SW pressures, when GMCs prevail in the dawn and prenoon sectors, the inflection of the envelope boundary is steeper, i.e. larger. At the same time, in the dawn and dusk sectors the inflection point $Bz_0$ shifts toward larger negative values (see Figure 8b). That result in faster IMF *Bz* influence saturation (at $Bz$~-15 nT) in the dawn sector, where the SW pressure required for GMCs is smaller, than that in the dusk sector, where the saturation occurs at $Bz$ < -20 nT and higher SW pressures. That is in good agreement with results obtained by *Yang et al.* [2003]. They predict that the threshold for saturation increases with SW dynamic pressure.

*3.2. Latitudinal dependence*

Latitudinal dependence of the lower boundary is studied in vicinity of noon (9 to 15 MLT) because of poor statistics at flanks (see Figure 2). In aGSM, the location of geosynchronous satellites is restricted by latitudes of about ±30° and, hence, at large MLT displacements from noon (say more than 3 hours or >45°) the effect of latitude diminishes in comparison with the longitudinal effect. The ranges of aGSM latitudes (*mLat*) used for determination of the envelope boundary are listed in Table 2. Here we assume that the magnetopause is symmetrical relative to the GSM equatorial plane. Actually that is not the case for large tilt angles. However, in the first approach we neglect this effect.

From Table 2 one can see that the SW pressure required for GMC increases with latitude. The accuracy of envelope boundary determination is quite high (correlation coefficients r>0.75). The latitudinal dependence of asymptotic pressures can be described well by a power function of sin(*mLat*) as shown in Figure 9:

$$dP_{max}(mLat) \equiv P_{max}(mLat) - 21.0 = 50.0 \cdot \sin^{1.66}(mLat) \qquad (6a)$$

$$dP_{min}(mLat) \equiv P_{min}(mLat) - 4.8 = 20.6 \cdot \sin^{1.92}(mLat) \qquad (6b)$$

Here we fit the residual $dP_{max}$ and $dP_{min}$, which are obtained after subtraction of the asymptotic pressures $P_{min}$ = 4.8 nPa and $P_{max}$ = 21. nPa derived for the lower boundary (Eq. 1). We have to point out that the power index of sin(*mLat*) is less than 2 due to blunted shape of the magnetopause. For the spherical shape, the exponent is expected to be equal to 2.

Because of insufficient statistics at middle GSM latitudes, we will not model the latitudinal dependencies for the steepness $\chi$ and inflection point $Bz_0$. We can only indicate that variation of those parameters with latitude is relatively small in comparison with the longitudinal dependence.

*3.3. Dependence on Dst*

Studying variations of the envelope boundary with geomagnetic parameters, we have found a strong dependence on *Dst*. In the present study we use *SYM-H* index as 1-min equivalent of hourly *Dst* index. Actually, most of the geosynchronous magnetopause crossings are observed during sudden commencement (SSC) or main phase of severe and strong magnetic storms. As one can see in Figure 10, the amplitude of SSC can reach up to >100 nT and the intensity of storms can be higher than -300 nT. The magnetic storms are accompanied by intensification of whole magnetospheric current system, including mainly ring current and cross-tail current. It is important to point out that the *Dst* variation has a relationship with *Bz*. However, this relationship is very complex and non-linear in time [e.g. *Burton et al.*, 1975; *O'Brien and McPherron*, 2002; *Wang, et al.*, 2003; *Siscoe et al.,* 2005; *Vasyliunas*, 2006] that results in very week linear correlation ($r = 0.17$) between *SYM-H* and *Bz*. Hence, *SYM-H* and *Bz* can be treated statistically as independent variables.

We study the dependence on *Dst* in 2-hour vicinity of minimum of SW pressure at ~11 MLT (see Figure 3), i.e. from 9 to 13 MLT. In Table 3 we list the ranges of *SYM-H* index for which we determine the envelope boundaries. An example of the boundary determination for *SYM-H* > -100 nT is presented in Figure 11. The envelope boundary is located above the lower boundary derived for *Dst* > -300 nT both for positive and negative *Bz*. In Figure 11 we also plot envelope boundaries obtained for other ranges of *Dst*. Note that the accuracy of the boundary determination for positive *Dst* is quite low (see Table 3) because of very low statistics at such conditions. However, the general tendency is supported by accurate determination (correlation coefficient r>0.7) of the envelope boundaries for negative *Dst*.

The dependence of asymptotic pressures $P_{max}$ and $P_{min}$ on the *Dst* variation can be approximated by an exponential function (see Figure 12):

$$dP_{max}(Dst) \equiv P_{max}(Dst) - 21. = 6.99 \cdot \exp(Dst/105.) \quad (7a)$$

$$dP_{min}(Dst) \equiv P_{min}(Dst) - 4.8 = 8.69 \cdot \exp(Dst/57.4) \quad (7b)$$

Hence, the asymptotic pressures decrease exponentially with increasing storm disturbances. It is important to note that for positive *Dst*, the $P_{max}$ is approaching to ~35 nPa.

Figure 13 shows a dependence of steepness $\chi$ on the *Dst* variation. In the range of *Dst* > -150, the steepness decreases linearly with increasing *Dst*. It seems that for large negative *Dst* < -200 nT this dependence is broken and the steepness does not change much. Hence, we can describe the dependence of steepness $\chi$ on the *Dst* by the following expressions:

$$\chi(Dst) = 0.0739 - Dst/1300 \ (Dst > \text{-150 nT})$$

(8)

$$\chi(Dst) = 0.2 \ (Dst < \text{-150 nT})$$

We have to point out that for positive *Dst* the steepness approaches to zero and, thus, the hyperbolic tangent function (see Eq. 2) approaches to a linear dependence of *Psw* on *Bz*. Such linear dependence was used in a number of magnetopause models developed for moderately disturbed conditions [e.g. *Petrinec and Russell*, 1996; *Shue et al.*, 1997].

From the above one can see that with decreasing geomagnetic activity, when the *Dst* is growing from negative to positive values, both maximal and minimal asymptotic pressures increases but the ratio of $P_{max}$ to $P_{min}$ as well as the steepness $\chi$ are decreasing. Note that the inflection points $Bz_0$ of the envelope boundaries for various *Dst* groups about 0 nT. Such behavior might indicate that the effectiveness of the magnetopause erosion under southward IMF increases during higher storm activity. In addition, the effect of southward IMF influence saturation is more prominent for large negative *Dst* and it vanishes for positive *Dst*.

*3.4. A predictive model*

Finally we can build a predictive model in the form of Equation 2 with the following coefficients:

$$P_{max} = P_{max}(mLon) + dP_{max}(mLat) + dP_{max}(Dst)$$

$$P_{min} = P_{min}(mLon) + dP_{min}(mLat) + dP_{min}(Dst)$$

(9)

$$Bz_0 = Bz_0(mLon)$$

$$\chi = \sqrt{\chi(mLon) \cdot \chi(Dst)}$$

Here $P_{max}(mLon)$, $dP_{max}(mLat)$, $dP_{max}(Dst)$, $P_{min}(mLon)$, $dP_{min}(mLat)$, $dP_{min}(Dst)$ are defined, respectively, by Equations 3a, 6a, 7a, 3b, 6b, 7b; and $Bz_0(mLon)$, $\chi(mLon)$, $\chi(Dst)$ are presented, respectively, by Equations 4, 5, 8. Note that the dependence on latitude is modeled in the range from 9 to 15 MLT. Outside this interval this dependence is diminished and only longitudinal and *Dst* effects persist. The dependence on *Dst*, derived in vicinity of noon, is expanded to the whole dayside magnetopause. The model allows predicting the SW pressure *Psw* required for GMC at given location (*mLon*, *mLat*), for given IMF *Bz* and geomagnetic *Dst* index. If the actual SW pressure is equal to or higher than the predicted one, then the magnetopause should cross the geosynchronous orbit at the given location.

Due to the limited statistics, we do not study dependencies of $Bz_0$ from latitude as well as how the dependence of *Dst* varies with MLT. The latter is most intrigues because it can show us how the IMF *Bz* saturation and asymptotic pressures change with the SW pressure. That will be a subject of further studies based on extended set of GMCs. In the present shape, the model can be considered as a first step in modeling the GMCs with taking into account the effects of dawn-dusk asymmetry and IMF *Bz* saturation.

## 4. Comparison with other models

In order to estimate the accuracy of the predictive model and compare it with other magnetopause models we use an extended data set, which includes both magnetosheath and magnetosphere geosynchronous intervals accumulated in 1995 to 2001 [*Suvorova et al.*, 2005]. This set consists of 5855 magnetosheath points and 9605 points collected inside the magnetosphere in vicinity of inbound and outbound magnetopause crossings. All the points are provided by the upstream solar wind conditions. It is important to note that the model was developed on the base of a portion of magnetosheath points. The points in the magnetosphere were not used in the modeling. Hence, the total data set of magnetosheath and magnetosphere intervals is practically independent from the data set used for the model construction and, thus, we can apply this data set for comparison of different models.

The estimation of accuracy is based on such statistical quantities as overestimation/underestimation ratio (*OUR*), probability of correct prediction (*PCP*), probability of detection (*PoD*) and false alarm rate (*FAR*) [e.g. *Shue et al.*, 2000; *Yang et al.*, 2002; *Dmitriev et al.*, 2003]. Those quantities are derived from four numbers *A*, *B*, *C*, and *D*, which are calculated in the following manner. The number *A* is a number of cases when a model correctly predicts that the magnetopause is located inside the geosynchronous orbit. The number *B* (*C*) is a number of wrong predictions, when a model underestimates (overestimates) the magnetopause distance. The quantity *D* is a number of correct rejections, when the model correctly predicts that the magnetopause is located outside the geosynchronous orbit. The sum of these four numbers gives the total number of points *N*. The statistical quantities *OUR*, *PCP*, *PoD* and *FAR* are defined as the following:

$$OUR = (C - B)/(C + B) \qquad (10a)$$

$$PCP = (A + D)/N \qquad (10b)$$

$$PoD = A/(A + C) \qquad (10c)$$

$$FAR = B/(B + A) \qquad (10.d)$$

The best model prediction should have *OUR* approaching to 0, highest *PCP* and *PoD*, and lowest *FAR*.

In Table 4 we compare our model with four magnetopause models developed for disturbed conditions. All those models enable prediction of the IMF $B_z$ saturation but in different manner. Models by *Shue et al.* [1998] and *Yang et al.* [2003] are axially symmetric. *Kuznetsov and Suvorova*'s [1998a] model predicts ~0.5 to 2 Re duskward shifting of the magnetopause under strong southward IMF. *Lin's et al.* [2010] model is a modern 3-D magnetopause model, which enables to describe magnetopause indentation in the cusp regions and tilt angle effect. However, this model does not describe dawn-dusk asymmetry.

We can see that the most sophisticated model by *Lin et al.* [2010] predicts quite well the magnetopause dynamics about geosynchronous orbit. Among the previous models, this model has highest *PCP* and *PoD*, and relatively low *OUR* and *FAR*. We have to note that the quantities *PoD* and *FAR* are not independent (see Equations 10c and 10d): they both depend on hit number *A*. A model, systematically overestimating the magnetopause distance (large *OUR*), has larger number *C* and smaller number *B*, that results in lower *FAR* and also lower *PoD*. This situation one can find for the models by *Shue et al.* [1998] and *Yang et al.* [2003], which are characterized by relatively high OUR (>0.5). Note that these two models are not very complex and have only 9 free parameters. The models by *Kuznetsov and Suvorova* [1998a] and by *Lin et al.* [2010] have relatively low *OUR*, and quite high *PCP* and *PoD*, though the *FAR* of those models is also high. Our predictive model is characterized by the lowest *OUR*, highest *PCP* and *PoD*, and relatively low *FAR*. i.e. it is able to predict both magnetosheath and magnetospheric intervals with practically equal success.

We have to note that the predictive model and the *Lin's et al.* [2010] model are characterized by similar complexity. The predictive model depends on three parameters: $P_{sw}$, $B_z$ and $D_{st}$. The model dependencies are fitted by 22 free parameters, which are required to describe four effects: latitudinal dependence, dawn-dusk asymmetry, IMF *Bz* influence saturation, and dependence on *Dst*. The model by *Lin et al.* [2010] depends also on three physical parameters: sum of solar wind dynamic and magnetic pressures, IMF *Bz*, and tilt angle of geodypole axis. The model uses 21 fitting parameters in order to describe four effects: dependence on solar wind pressure, dependence on IMF *Bz* with saturation, north-south asymmetry, and magnetopause indentation in the cusp regions. We can see that two models describe the same number of effects, depend on the same number of physical parameters and are fitted by similar number of free parameters. Hence, higher score of the predictive model means that this model is indeed better for prediction of the geosynchronous magnetopause crossings.

We have to point out that our model can not be converted into the traditional shape of magnetopause models predicting the magnetopause distance as a function of upstream conditions. However, we can compare some asymptotical parameters. In the modern models [*Shue et al.*, 1998; *Lin et al.*, 2010], the subsolar magnetopause distance $r_0$ is expressed in the following manner:

$$r_0 = \{R_0 + a \tanh[k (Bz+Bz_0)]\} Psw^\gamma \quad (11)$$

From the predictive model we can calculate parameters $R_0$ and *a*. It is easy to show that in asymptotic approach, when IMF $Bz \rightarrow -\infty$ or $+\infty$, Equation 11 is converted into the following expressions:

$$R_0 = \frac{r_0}{Pd_{min}^\gamma} \quad \text{(for Bz} \rightarrow -\infty) \quad (12a)$$

$$a = \frac{r_0}{Pd_{max}^\gamma} - \frac{r_0}{Pd_{min}^\gamma} \quad \text{(for Bz} \rightarrow +\infty) \quad (12b)$$

The model by *Shue et al.* [1998] gives $\gamma = -0.152$, $R_0 = 10.22$ Re, $a = 1.29$ Re, and the model by *Lin et al.* [2010] gives $\gamma = -0.194$, $R_0 = 12.5$ Re, $a = 3.81$ Re. In our case of geosynchronous orbit ($r_0 = 6.6$ Re), we can estimate for non-storm conditions ($Dst > 0$, $Pd_{min} = 19$ nPa and $Pd_{max} = 35$ nPa) that $R_0 = 10.3$ Re and $a \sim 1.0$ for the *Shue et al.* [1998] model and $R_0 = 11.7$ Re and $a \sim 1.47$ for the *Lin et al.* [2010] model. These values vary with geomagnetic activity. For strong magnetic storms ($Dst < -200$ nT, $Pd_{min} = 4.8$ nPa and $Pd_{max} = 21$ nPa) we obtain $R_0 = 8.4$ Re and $a \sim 2.1$ Re for the *Shue et al.* [1998] model and $R_0 = 8.95$ Re and $a \sim 3$ Re for the *Lin et al.* [2010] model. One can see that the values of $R_0$ and $a$, derived from the predictive model, do not contradict to the numbers obtained in the previous models on the base of fitting the magnetopause crossings at various distances.

## 5. Discussion

In contrast to the previous magnetopause models, the predictive model does not calculate the magnetopause distance but shows whether or not the magnetopause crosses the geosynchronous orbit at given location and for given solar wind and geomagnetic conditions. The model provides highest *PCP* and *PoD* quantities. The *OUR* is very close to 0. It means that the model is well balanced, i.e. overestimation and underestimation scores are equal and, hence, the model disadvantages are caused rather by a noise than by systematical errors in modeling of control parameters. The noise of the model is not very low that results in relatively high FAR. This noise is originated from the effects and dependencies, which we neglect or do not take into account.

Using the 3-D magnetopause model by *Lin et al.* [2010] we can estimate that ~30° variation of the tilt angle changes the location of subsolar magnetopause at geosynchronous orbit only by 0.2 Re. Hence, in the first approach we can neglect the effect of tilt angle. Perhaps, more important contribution to the noise is produced by the MLT dependence of *Dst* effect. Because of limited statistics, in the present study we restrict the modeling of this effect by the range of 9 to 13 MLT. In Figures 11 and 12 one can clearly see a significant contribution of the *Dst* to the magnetopause location at geosynchronous orbit. The SW pressure required for GMCs decreases exponentially with increasing negative *Dst* variation. That is equivalent to substantial inward motion of the dayside magnetopause. At larger GSM longitudes, where higher pressures are required for GMCs, the dependence represented by Equations 7 and 8 can be considerably differ. However, more statistics are required for modeling this effect. Note that the dependence on MLT is strongly asymmetrical (see Figure 7).

The storm-time negative *Dst* variation is contributed by two magnetospheric currents: ring current and cross-tail current. During strong geomagnetic storms the magnetic effect of the tail current to the dayside geomagnetic field can be quite large and comparable with magnetic depletion produced by the ring current [*Maltsev et al.*, 1996; *Turner et al.*, 2000; *Alekseev et al.*, 2001]. While the magnetic effect of the ring current to the magnetopause is still controversial, the depletion of the dayside geomagnetic field by the cross-tail current is well established. This depletion results from the storm-time intensification of the cross-tail current and from sunward motion of its inner edge such that the enhanced current approaches to the dayside magnetopause. The negative magnetic effect of the cross-tail current increases with storm activity. Hence, we can attribute the decrease of SW pressure required for GMCs to the enhancement of dayside magnetic field depletion produced by the storm-time cross-tail current.

Another important effect, related to strong southward IMF and large negative *Dst*, is saturation of the IMF *Bz* influence to the magnetopause. In Figure 13 we demonstrate that the steepness $\chi$ of the envelope boundary increases with decreasing *Dst*. It means that for large negative *Dst* the threshold for saturation moves toward smaller magnitudes of negative *Bz*. This effect we can also find in dynamics of the envelope boundaries with *Dst* presented in Figure 11.

There has been no complete physical explanation proposed both for the *Bz*-influence saturation and for the dawn-dusk asymmetry. Various mechanisms have been proposed to explain the magnetopause dawn-dusk asymmetry. One of them is a predominant IMF orientation along the Parker spiral [e.g. *Russell et al.*, 1997]. On the other hand, *Burlaga et al.* [1987] demonstrated strong variations of the IMF vector orientation in compound streams and magnetic clouds. *Smith and Phillips* [1997] concluded that coronal mass ejections (CME), the shocks upstream of CMEs, and other interplanetary shocks are responsible for the apparent deviation of the IMF spiral relative to the Parker prediction. Comprehensive statistical analyses [*Dmitriev et al.*, 2009; *Borovsky*, 2010] convincingly show an existence of isotropic population of the IMF direction vectors which is contributed by interplanetary shocks, ejecta and CIR crossings. We have to point out that geosynchronous magnetopause crossing occur during very strong interplanetary disturbances, associated with after-shock sheath regions and CMEs. Those interplanetary structures are characterized by non-Parker random IMF orientation [e.g. *Dmitriev et al.*, 2004].

One of the most probable sources of the dawn-dusk asymmetry is the asymmetric storm-time ring current, with a maximum in the evening sector developing under strong southward IMF *Bz* [*McComas et al.*, 1993; *Itoh and Araki,* 1996; *Dmitriev et al.*, 2004; 2005]. Due to this asymmetric ring current, the dusk side of the magnetosphere, where the ring current is maximal, should be larger than the dawn side [*Cummings,* 1966; *Burton et al.,* 1975]. However, magnetic effect of the ring current to the magnetopause is still a subject of discussions.

Besides the magnetic effect, there is also a thermal pressure of the magnetospheric plasma $P_{tm}$, which is dominated by ring current particles. Direct measurements of the thermal plasma in the magnetosphere [*Frank,* 1967; *Lui et al.,* 1987; *Lui and Hamilton*, 1992] show that the perpendicular pressure in the dayside region of geosynchronous orbit is about 1~2 nPa for quiet geomagnetic conditions and grows up to 4 nPa during strong geomagnetic storms. This pressure is comparable with the SW pressure in the "regime of saturation", $Psw$=4.8 nPa. Therefore, for large negative IMF *Bz* a contribution of the magnetospheric thermal pressure to the pressure balance at the magnetopause can not be neglected. As a result we can amend the pressure balance equation in the nose region:

$$k \cdot Psw = \frac{(2fB)^2}{8\pi} + P_{tm} \qquad (13)$$

This suggestion is supported by results of high-resolution 3-D MHD simulations reported by *Borovsky et al.*, [2008]. They found that the reconnection can be saturated by a ''plasmaspheric effect'': high-density magnetospheric plasma flows from plasmasphere into the magnetopause reconnection site and mass loads the reconnection such the reconnection rate is reduced.

The compression and erosion affect the magnetic field at the magnetopause in a different manner, so there are observed two different types of GMC events based on the morphology of the magnetic field signatures [*Rufenach et al.,* 1989; *Itoh and Araki,* 1996]. The magnetic effect of geomagnetic currents to the subsolar magnetic field can modify the coefficient *f*. Namely, the *f* decreases due to the depletion of dayside magnetic field by the cross-tail and field-aligned currents. Using this fact *Kuznetsov and Suvorova* [1998b] further concluded that the erosion on the dayside magnetopause under strong negative $B_z$ is accompanied by a decrease of the coefficient *f* from 1 down to 0.5. *Ober et al.* [2006] reported similar decrease of the *f* due to the magnetic effect of field-aligned currents. In other words, the contribution of geomagnetic field pressure to the pressure balance decreases and the relative importance of the thermal pressure $P_{th}$ grows up.

Under strong negative $B_z$, the magnetopause moves earthward due to reconnection, which leads to penetration of the IMF to smaller distances. However, the IMF influence can be terminated by a force of non-magnetic nature such as thermal pressure of the magnetospheric plasma. Hence, we can suggest that the "$B_z$-influence saturation" might be caused by the enhanced contribution of the magnetospheric thermal pressure to the pressure balance at the dayside magnetopause during strong magnetic storms.

We have to point out that the magnetopause nose point, where the pressure balance can be represented simply by Equation 13, does not coincide with the subsolar point because of duskward skewing of the magnetopause. As a result, the perigee point, where the magnetopause approaches mostly close to geosynchronous orbit, is shifted toward the dawn while the nose point is shifted toward the dusk and, hence, located at a larger geocentric distance. In other words, the SW conditions for a GMC at the nose point should be stronger than the minimum necessary solar wind conditions for a GMC. This difference depends on the magnetopause shift *dY*, which is related to the ring current asymmetry, and on the magnetopause flaring, which depends on the *Bz*.

Using the asymmetrical model by *Kuznetsov and Suvorova* [1998a] we can roughly estimate that for *Bz*=-30 nT and ~2 Re duskward shift of the magnetopause, the SW pressure required to push the magnetopause nose point into a distance of 6.6 Re should be larger by 50% than the minimum necessary SW pressure for a GMC at the perigee point (*Psw* = 4.8 nPa). Hence a GMC at the MP nose point requires *Psw* ~ 7 nPa for large negative *Bz*. In Table 1 we can find similar values of $P_{min}$ in postnoon sector. Such SW pressure is in good agreement with our suggestion regarding the magnetospheric plasma pressure contribution to the pressure balance at the MP. Indeed, estimation of the geodipole magnetic field energy density (assuming *f*=0.5) at geosynchronous orbit gives us a value for the magnetic field pressure of about 4.6 nPa (the first term in Equation 13). Our estimation of the SW dynamic pressure, required for a GMC at the nose point, gives *Psw* = 7 nPa. The difference of about 2 nPa can be attributed to the thermal pressure of the magnetospheric plasma (the second term in Equation 13), which is in good agreement with experimental measurements [*Lui et al.,* 1987; *Liemohn et al.,* 2008].

Finally, we have to point out that using the *Dst* index in modeling the magnetopause allows taking into account a non-linear integral dependence from the interplanetary and magnetospheric conditions, i.e. so-called "effect of prehistory". It was established that the *Dst* variation is a time integral of the interplanetary induced electric field [*Burton et al.*, 1975]. *Burke et al.* [2007] show that the *Dst* variation correlates very well with an integral of the temporal variation of polar cap potential divided by the width of the magnetosphere. Hence, the *Dst* index accumulates such effects as preconditioning by IMF *Bz* and magnetospheric prehistory.

## 6. Conclusions

1. A predictive model of geosynchronous magnetopause crossings has been developed on the base of large statistics of magnetosheath interval detected by geosynchronous satellites GOES and LANL in 1995 to 2001.

2. The model describes such nonlinear effects as dawn-dusk magnetopause asymmetry, IMF $Bz$ influence saturation and effects of preconditioning by IMF $Bz$ and magnetospheric prehistory.

3. In statistical comparison with other models, the predictive model demonstrates the highest score: best capability for GMC prediction (highest PCP, PoD, OUR ~ 0) and very low false alarm rate.

4. We have found a strong decrease of the solar wind pressure required for GMCs with increasing negative $Dst$ variation. This effect can be attributed to a depletion of the dayside magnetic field by the cross-tail current intensified during magnetic storms.

5. Diminishing of the IMF $Bz$ saturation effect for small negative and positive $Dst$ is a strong support of the suggestion that the enhanced thermal pressure of the magnetospheric plasma and ring current particles is responsible for the saturation of magnetic effect of the IMF $Bz$ at the dayside magnetopause.

**Acknowledgements** This work was supported by grant NSC-98-2111-M-008-019 from the National Science Council of Taiwan and by Ministry of Education under the Aim for Top University program at National Central University of Taiwan #985603-20.

**References**

Alexeev, I. I., V. V. Kalegaev, E. S. Belenkaya, S. Y. Bobrovnikov, Ya. I. Feldstein, and L. I. Gromova (2001), Dynamic model of the magnetosphere: Case study for January 9-12, 1997, *J. Geophys. Res., 106*, 25,683.

Borovsky, J. E., M. Hesse, J. Birn, and M. M. Kuznetsova (2008), What determines the reconnection rate at the dayside magnetosphere? *J. Geophys. Res.*, 113, A07210, doi:10.1029/2007JA012645.

Borovsky, J. E. (2010), On the variations of the solar wind magnetic field about the Parker spiral direction, *J. Geophys. Res.*, 115, A09101, doi:10.1029/2009JA015040.

Burke, W. J., L. C. Gentile, and C. Y. Huang (2007), Penetration electric fields driving main phase Dst, *J. Geophys. Res., 112*, A07208, doi:10.1029/2006JA012137.

Burlaga, L., K. Behannon, and L. Klein (1987), Compound Streams, Magnetic Clouds, and Major Geomagnetic Storms, *J. Geophys. Res.*, 92(A6), 5725-5734.

Burton, R. K., R. L. McPherron, and C. T. Russell (1975), An empirical relationship between interplanetary conditions and *Dst*, *J. Geophys. Res., 80*, 4204.

Collier, M. R., J. A. Slavin, R. P. Lepping, A. Szabo, and K. Ogilvie (1998), Timing accuracy for the simple planar propagation of magnetic field structures in the solar wind, *Geophys. Res. Lett., 25*, 2509.

Cummings, W.D. (1966), Asymmetric ring current and the low-latitude disturbance daily variation. *J. Geophys. Res., 71*, 4495.

Dmitriev, A.V., and A.V. Suvorova (2000), Three-dimensional artificial neural network model of the dayside magnetopause, *J. Geophys. Res., 105*, 18909.

Dmitriev A., J.-K. Chao, D.-J. Wu (2003), Comparative study of bow shock models using Wind and Geotail observations, *J. Geophys. Res., 108*(A12), 1464, doi:10.1029/2003JA010027.

Dmitriev A.V., A.V. Suvorova, J.-K. Chao, Y.-H. Yang (2004), Dawn-dusk asymmetry of geosynchronous magnetopause crossings, *J. Geophys. Res., 109*, A05203 doi: 10.1029/2003JA010171.

Dmitriev A., J.-K. Chao, M. Thomsen, A. Suvorova (2005), Geosynchronous magnetopause crossings on October 29-31, 2003, *J. Geophys. Res.*, *110*(A8), A08209, doi:10.1029/2004JA010582.

Dmitriev, A.V., A.V. Suvorova, I.S. Veselovsky (2009), Statistical Characteristics of the Heliospheric Plasma and Magnetic Field at the Earth's Orbit during Four Solar Cycles 20-23, in Handbook on Solar Wind: Effects, Dynamics and Interactions, Ed. Hans E. Johannson, NOVA Science Publishers, Inc., New York, 2009, p. 81-144.

Frank, L.A. (1967), On the extraterrestrial ring current during geomagnetic storms, *J. Geophys. Res.*, *72*, 3753.

Itoh K., and T. Araki (1996), Analysis of geosynchronous magnetopause crossings, in *Proceedings of Solar Terrestrial Predictions Workshop*, Hitachi, Japan, Jan.23-27, 26.

Kuznetsov S.N., and A.V. Suvorova (1994), Influence of solar wind to some magnetospheric characteristics, in *Proceedings of WDS'94, PartII-Physics of plasmas and ionized media*, edited by J.Safrankova, pp.116-123, Charles University, Prague.

Kuznetsov S.N., and A.V. Suvorova (1996), On two regimes of solar wind interaction with magnetosphere, paper presented at First EGS Alfven Conference on Low-Altitude Investigation of Dayside Magnetospheric Boundary Processes, Kiruna, Sweden, Sep.9-13. 1996.

Kuznetsov, S.N., and A.V. Suvorova (1997), Magnetopause shape near geosynchronous orbit (in Russian), *Geomagn. Aeron.*, *37*, 1.

Kuznetsov, S.N., and A.V. Suvorova (1998a), An empirical model of the magnetopause for broad ranges of solar wind pressure and Bz IMF, in *Polar cap boundary phenomena*, *NATO ASI Ser.*, edited by J.Moen, A.Egeland and M.Lockwood, pp.51-61, Kluwer Acad., Norwell, Mass..

Kuznetsov, S. N., and A. V. Suvorova (1998b), Solar wind magnetic field and plasma during magnetopause crossings at geosynchronous orbit, *Adv. Space Res., 22*(1), 63.

Liemohn, M. W., J.-C. Zhang, M. F. Thomsen, J. E. Borovsky, J. U. Kozyra, and R. Ilie (2008), Plasma properties of superstorms at geosynchronous orbit: How different are they?, *Geophys. Res. Lett.*, 35, L06S06, doi:10.1029/2007GL031717.

Li, H., C. Wang, and J. R. Kan (2010), Midday magnetopause shifts earthward of geosynchronous orbit during geomagnetic superstorms with Dst $\leqslant$ 300 nT, *J. Geophys. Res.*, *115*, A08230, doi:10.1029/2009JA014612.

Lin, R. L., X. X. Zhang, S. Q. Liu, Y. L. Wang, and J. C. Gong (2010), A three-dimensional asymmetric magnetopause model, *J. Geophys. Res.*, *115*, A04207, doi:10.1029/2009JA014235.

Lui, A.T.Y., and D.C. Hamilton (1992), Radial profiles of quiet time magnetospheric parameters, *J. Geophys. Res., 97*, 19,325.

Lui, A.T.Y., R.W. McEntrie, and S.M. Krimings (1987), Evolution of the ring current during two geomagnetic storms, *J. Geophys. Res., 92*, 7459.

McComas, D. J., D. J., S. J. Bame, B. L. Barraclough, J. R. Donart, R. C. Elphic, J. T. Gosling, M. B. Moldwin, et al. (1993), Magnetospheric plasma analyzer (MPA): Initial three-spacecraft observations from geosynchronous orbit, *J. Geophys. Res.*, *98*, 13,453.

Maltsev, Y. P., A. A. Arykov, E. G. Belova, B. B. Gvozdevsky, and V. V. Safargaleev (1996), Magnetic flux redistribution in the storm time magnetosphere, *J. Geophys. Res.*, *101*, 7697.

Ober, D. M., M. F. Thomsen, and N. C. Maynard (2002), Observations of bow shock and magnetopause crossings from geosynchronous orbit on 31 March 2001, *107*(A8), 1206; doi:10.1029/2001JA000284.

Ober, D. M., N. C. Maynard, W. J. Burke, G. R. Wilson, and K. D. Siebert (2006), "Shoulders" on the high-latitude magnetopause: Polar/GOES observations, *J. Geophys. Res.*, *111*, A10213, doi:10.1029/2006JA011799.

O'Brien, T. P., and R. L. McPherron, Seasonal and diurnal variation of Dst dynamics, *J. Geophys. Res., 107*(A11), 1341, doi:10.1029/2002JA009435, 2002.

Odenwald, S. F., and J. L. Green (2007), Forecasting the impact of an 1859-caliber superstorm on geosynchronous Earth-orbiting satellites: Transponder resources, Space Weather, 5, S06002, doi:10.1029/2006SW000262.

Petrinec, S. M., and C. T. Russell (1996), Near-Earth magnetotail shape and size as determined from the magnetopause flaring angle, *J. Geophys. Res.*, *101*, 137.

Richardson, J. D., and K. Paularena (2001), Plasma and Magnetic Field Correlations in the Solar Wind, *J. Geophys. Res., 106*, 239.

Rufenach C.L., Jr.,R.F., Martin, and H.H. Sauer (1989), A study of geosynchronous magnetopause crossings, *J. Geophys. Res.*, *94*, 15,125.

Russell, C. T. (1976), On the occurrence of magnetopause crossings at 6.6 RE, *Geophys. Res. Lett.*, *3*(10), 593.

Russell, C. T., M. Ginskey, and S. M. Petrinec (1994a), Sudden impulses at low-latitude stations: Steady state response for northward interplanetary magnetic field, *J. Geophys. Res.*, *99*, 253.

Russell, C. T., M. Ginskey, and S. M. Petrinec (1994b), Sudden impulses at low-latitude stations: Steady state response for southward interplanetary magnetic field, *J. Geophys. Res.*, *99*, 13,403.

Russell, C. T., S. M. Petrinec, T. L. Zhang, P. Song, and H. Kawano, The effect of foreshock on the motion of the dayside magnetopause, *Geophys. Res. Lett.*, *24*, 1439, 1997.

Shue, J.-H., J. Chao, H. Fu, C. Russell, P. Song, K. Khurana, and H. Singer (1997), A new functional form to study the solar wind control of the magnetopause size and shape, *J. Geophys. Res.*, *102*(A5), 9497.

Shue, J.-H., P. Song, C.T. Russell, J.T. Steinberg, J.K. Chao et al. (1998), Magnetopause location under extreme solar wind conditions, *J. Geophys. Res.*, *103*, 17,691.

Shue, J.-H., P. Song, C. T. Russell, J. K. Chao, and Y.-H. Yang (2000), Toward predicting the position of the magnetopause within geosynchronous orbit, *J. Geophys. Res.*, *105*(A2), 2641.

Siscoe, G., J. Raeder, and A. J. Ridley (2004), Transpolar potential saturation models compared, *J. Geophys. Res.*, 109, A09203, doi:10.1029/2003JA010318.

Siscoe, G., R. L. McPherron, M. W. Liemohn, A. J. Ridley, and G. Lu (2005a), Reconciling prediction algorithms for *Dst*, *J. Geophys. Res.*, 110, A02215, doi:10.1029/2004JA010465.

Song, P., C. T. Russell, and M. F. Thomsen (1992), Slow Mode Transition in the Frontside Magnetosheath, *J. Geophys. Res.*, 97(A6), 8295–8305.

Smith, C., and J. Phillips (1997), The role of coronal mass ejections and interplanetary shocks in interplanetary magnetic field statistics and solar magnetic flux ejection, *J. Geophys. Res.*, 102(A1), 249-261.

Suvorova, A., A. Dmitriev, J.-K. Chao, Y.-H. Yang, M. Thomsen (2003), Necessary conditions for the geosynchronous magnetopause crossings, *Abstract on EGS-AGU-EUG Joint Assembly*, Nice, France, 7-11 April 2003.

Suvorova A., A. Dmitriev, J.-K. Chao, M. Thomsen, Y.-H. Yang (2005), Necessary conditions for the geosynchronous magnetopause crossings, *J. Geophys. Res.*, *110*, A01206, doi:10.1029/2003JA010079.

Turner, N.E., D.N. Baker, T.I. Pulkkinen, and R.L. McPherron (2000), Evaluation of the tail current contribution to Dst, *J. Geophys. Res., 105*, 5431.

Vasyliūnas, V. M. (2006), Reinterpreting the Burton-McPherron-Russell equation for predicting Dst, *J. Geophys. Res.*, *111*, A07S04, doi:10.1029/2005JA011440.

Weimer, D. R., D. M. Ober, N. C. Maynard, W. J. Burke, M. R. Collier (2002), Variable time delays in the propagation of the interplanetary magnetic field, *J. Geophys. Res.*, *107*(A8), 10.1029/2001JA009102.

Wrenn, G.L., J.F.E. Johnson, A.J. Norris, and M.F. Smith (1981), GEOS-2 magnetopause encounters: Low energy (<500 eV) particle measurements, *Adv. Space Res.*, *1*, 129.

Yang Y.-H., J. K. Chao, C.-H. Lin, and J.-H. Shue, X.-Y. Wang, et al. (2002), Comparison of three magnetopause prediction models under extreme solar wind conditions, *J. Geophys. Res.*, *107*, 10.1029, SMP 3-1.

Yang, Y.-H., J. K. Chao, A. V. Dmitriev, C.-H. Lin, and D. M. Ober (2003), Saturation of IMF Bz influence on the position of dayside magnetopause, *J. Geophys. Res.*, *108*(A3), 1104, doi:10.1029/2002JA009621.

**Table 1.** Parameters of the envelope boundary in various MLT ranges

| MLT range | $P_{min}$, nPa | $P_{max}$, nPa | r |
|:---:|:---:|:---:|:---:|
| 6 - 8 | 7.7 | 75 | .97 |
| 6 - 8.5 | 7 | 60 | .75 |
| 6 - 9 | 5.2 | 50 | .78 |
| 6 - 9.5 | 5.2 | 30 | .84 |
| 6 - 10 | 5 | 28 | .87 |
| 6 - 10.5 | 4.8 | 27 | .88 |
| 6 - 11 | 4.8 | 22 | .84 |
| 11.5 - 18 | 5 | 21 | .91 |
| 12 - 18 | 5.1 | 22 | .92 |
| 12.5 - 18 | 6.3 | 22 | .83 |
| 13 - 18 | 7 | 22 | .93 |
| 13.5 - 18 | 7 | 30 | .81 |
| 14 - 18 | 7.5 | 35 | .85 |
| 14.5 - 18 | 11 | 45 | .95 |
| 15 - 18 | 13 | 50 | .89 |
| 16 - 18 | 21 | 75 | .83 |

**Table 2.** Parameters of the envelope boundary for various aGSM latitudes

| $|mLat|$, deg | $P_{min}$, nPa | $P_{max}$, nPa | r |
|:---:|:---:|:---:|:---:|
| >0 | 4.8 | 21 | .91 |
| >5 | 5 | 22 | .89 |
| >10 | 5.5 | 23 | .85 |
| >15 | 6 | 27 | .76 |
| >20 | 8 | 30 | .83 |

**Table 3.** Parameters of the envelope boundary for various $Dst$

| $Dst$, nT | $P_{min}$, nPa | $P_{max}$, nPa | $P_{max}/P_{min}$ | r |
|:---:|:---:|:---:|:---:|:---:|
| >-400 | 4.8 | 21 | 4.4 | 0.91 |
| >-200 | 5 | 22 | 4.4 | 0.90 |
| >-150 | 5.6 | 23 | 4.1 | 0.86 |
| >-100 | 6 | 25 | 4.2 | 0.91 |
| >-70 | 7 | 27 | 3.9 | 0.88 |
| >-50 | 9 | 27 | 3.0 | 0.75 |
| >-30 | 11 | 30 | 2.7 | 0.93 |
| >0 | 13 | 33 | 2.5 | 0.68 |
| >20 | 19 | 35 | 1.8 | 0.08 |

**Table 4.** Comparison of the magnetopause models

| Model | OUR | PCP | POD | FAR |
|---|:---:|:---:|:---:|:---:|
| *Shue et al.* [1998] | .57 | .77 | .52 | .20 |
| *Yang et al.* [2003] | .59 | .77 | .52 | .19 |
| *Kuznetsov & Suvorova* [1998] | .15 | .78 | .66 | .27 |
| *Lin et al.* [2010] | .21 | .80 | .68 | .24 |
| Present model | .08 | .82 | .74 | .23 |

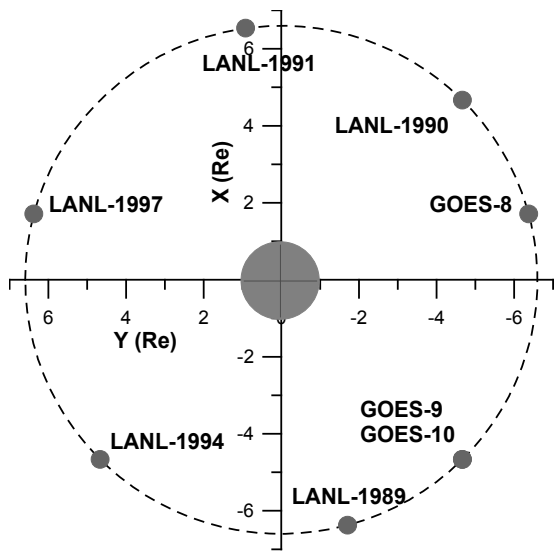

Figure 1. Geographyc location of GOES and LANL geosynchronous satellites.

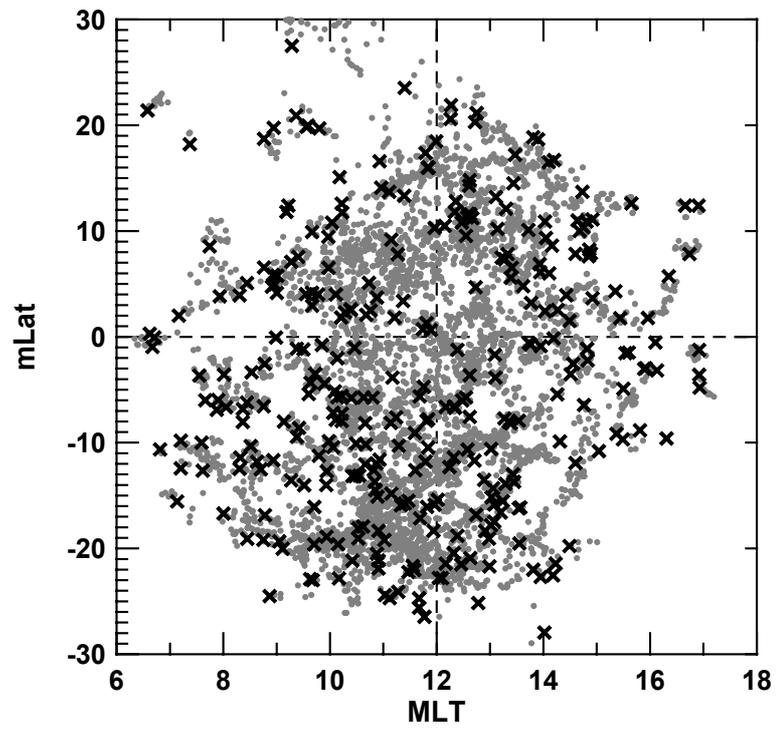

Figure 2. Scatter plot of the collected geosynchronous magnetopause crossings (GMCs) (depicted by black crosses) and magnetosheath (MSh) intervals (gray circles) in aberrated GSM (aGSM) coordinates latitude (*mLat*) versus magnetic local time (MLT).

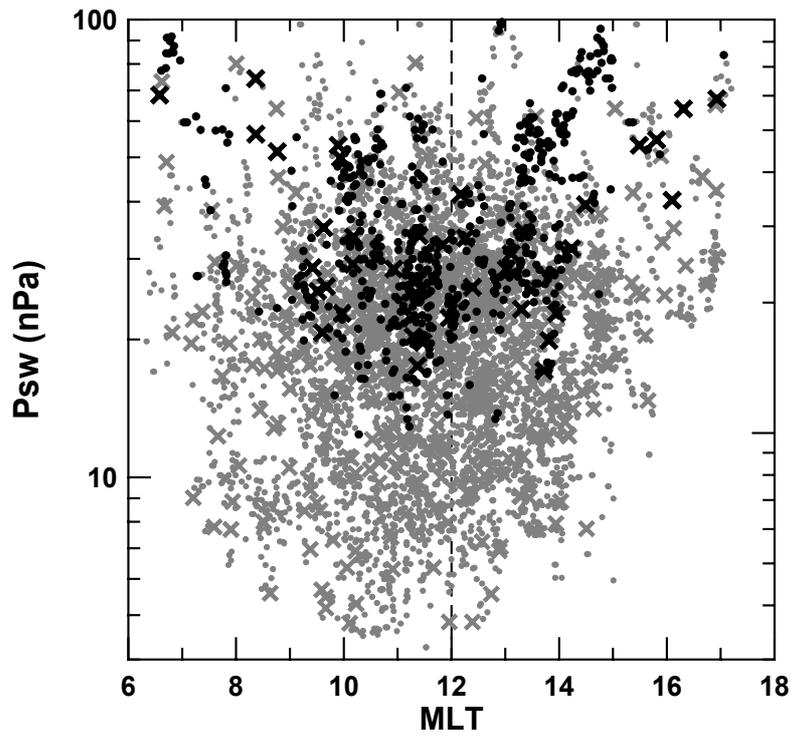

Figure 3. Scatter plot of total solar wind pressure *Psw* versus MLT for all collected GMCs (gray crosses) and MSh intervals (gray circles), and for those under *Bz* > 5 nT (black crosses and circles, respectively). A noticeable dawn-dusk asymmetry is clearly seen for whole statistics, while for *Bz* > 5 nT, the asymmetry diminishes.

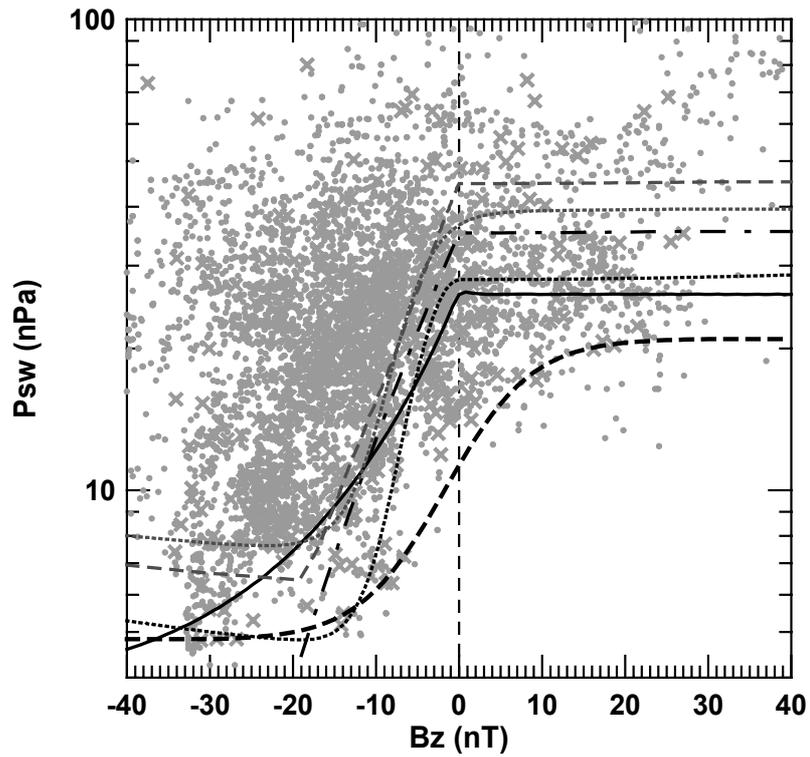

Figure 4. Scatter plot of total solar wind pressure *Psw* versus IMF *Bz* in aGSM for the collected GMCs (gray crosses) and MSh intervals (gray circles). Different model predictions of the solar wind conditions required for GMCs are shown by different curves: black dashed-dotted curve [*Petrinec and Russell*, 1996]; black dotted curve [*Kuznetsov & Suvorova*, 1998a]; gray dotted curve [*Shue et al.*, 1998]; gray dashed curve [*Yang et al.*, 2003]; black solid curve [*Lin et al.*, 2010]. The black dashed curve depicts a lowest envelope boundary (see Eq. 1) of the solar wind conditions required for GMCs [*Suvorova et al.*, 2005].

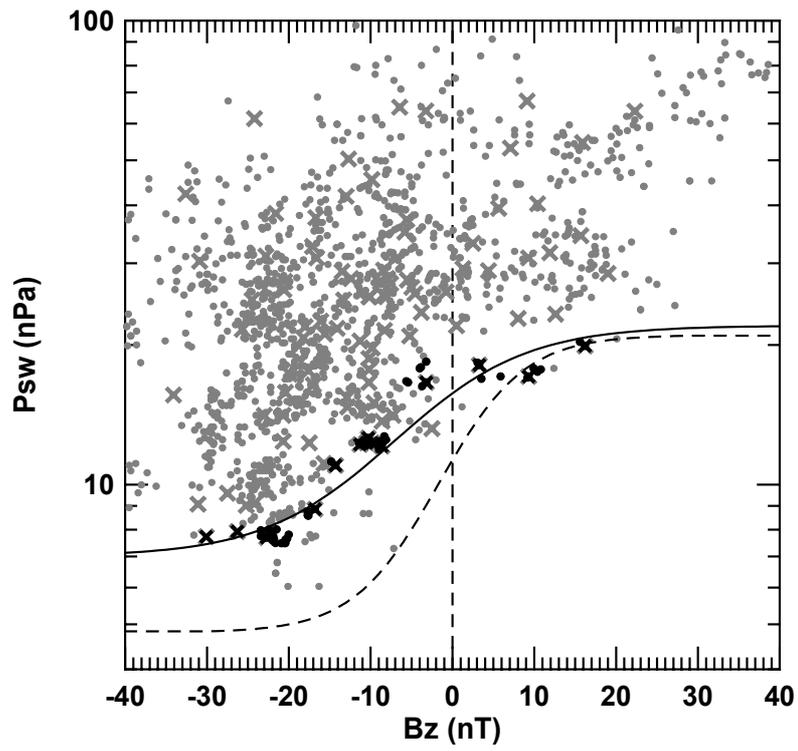

Figure 5. The same as in Figure 4 but for 13 - 18 MLT range. The best fit of envelope boundary is indicated by thick solid curve. The GMCs and MSh intervals selected for the boundary fitting are indicated by black symbols.

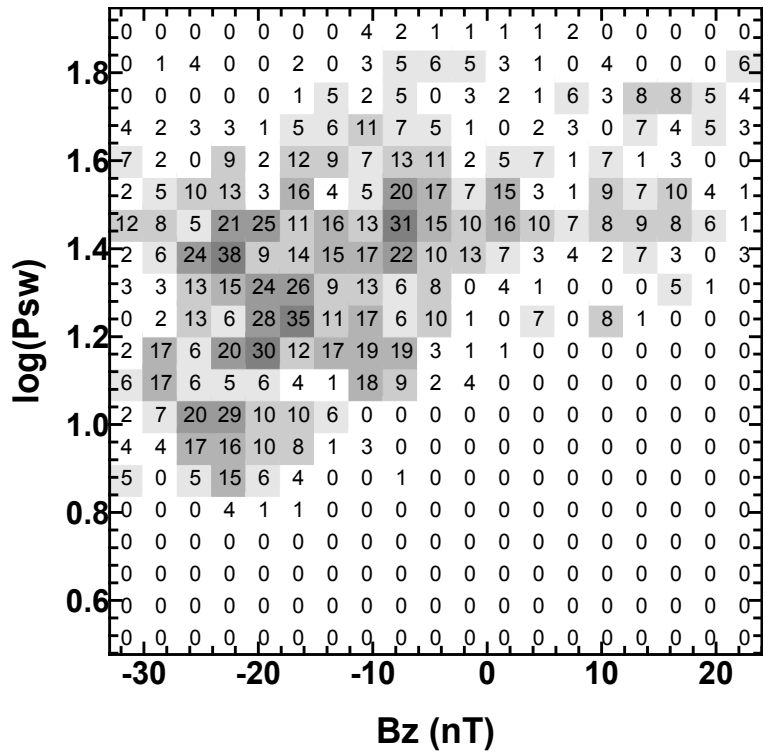

Figure 6. Two-dimensional distribution of occurrence number of the MSh intervals binned in the coordinates *Psw* (in logarithmic scale) versus IMF *Bz* in aGSM. The occurrence number is indicated for each bin and varies from <5 (white bins) to >30 (dark gray bins). The statistically significant gray bins with the lowest *Psw* for each given *Bz* indicate to approximate location of the envelope boundary, beyond which the occurrence number decreases sharply from ≥5 (gray bins) to <5 (white bins).

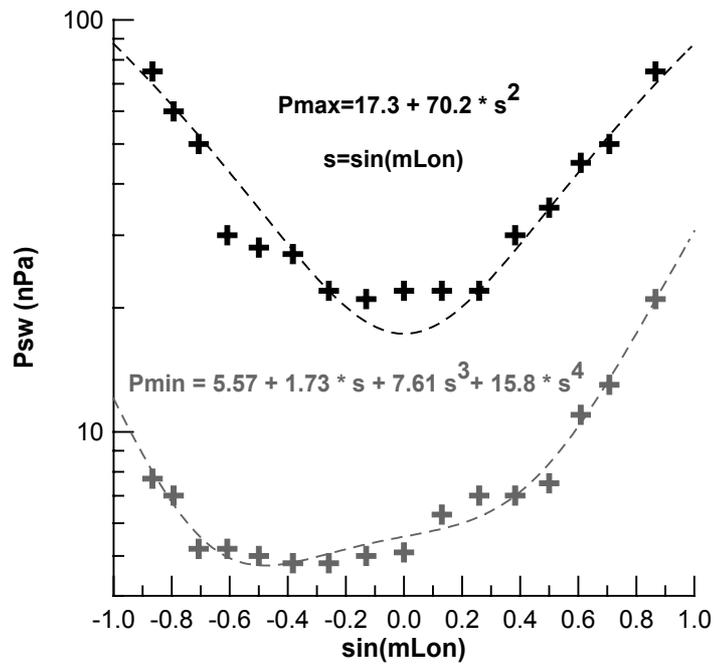

Figure 7. Maximal $P_{max}$ and minimal $P_{min}$ asymptotic pressures of the envelope boundary obtained in various ranges of aGSM longitude, *mLon*. The dashed lines correspond to best fit of the asymptotic pressures by a polynomial function of sin(*mLon*).

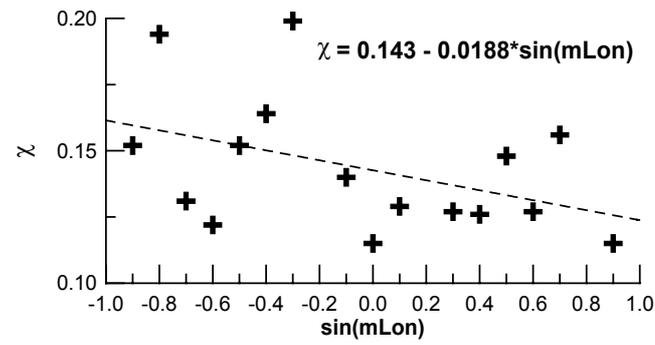

a

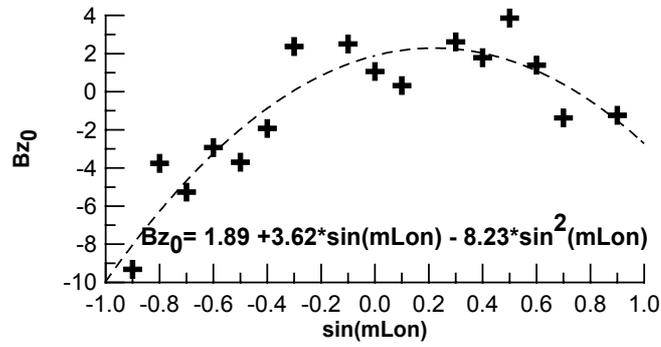

b

Figure 8. Parameters of the envelope boundary (a) $\chi$ and (b) $Bz_0$ calculated in various longitudinal ranges. The dashed lines correspond to best fit of these parameters by a polynomial function of sine of aGSM longitude.

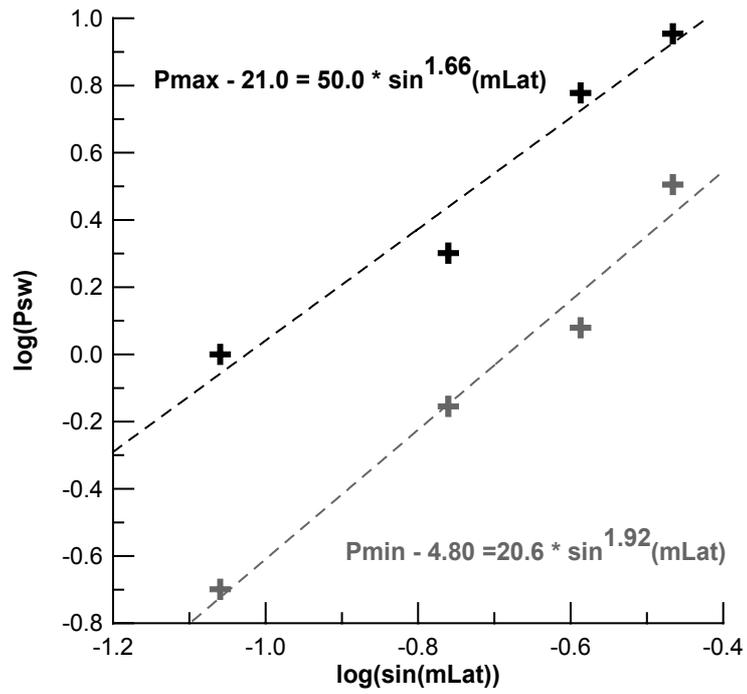

Figure 9. Maximal $P_{max}$ and minimal $P_{min}$ asymptotic pressures of the envelope boundary obtained in noon sector for various ranges of aGSM latitude, *mLat*. The dashed lines correspond to best fit of the pressures by a function of sin(*mLat*).

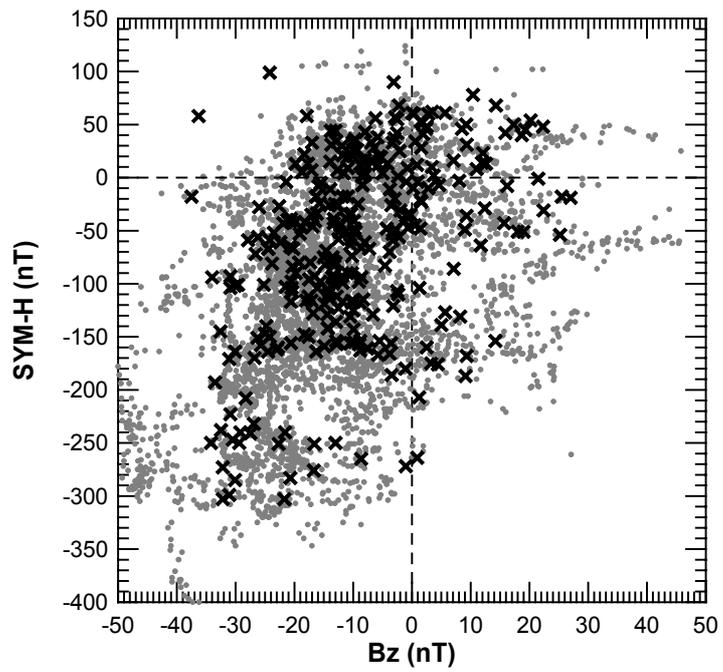

Figure 10. Scatter plot of *SYM-H* index versus IMF *Bz* in aGSM for the GMCs (black crosses) and MSh intervals (gray circles). Most of GMCs occur during severe and strong magnetic storms.

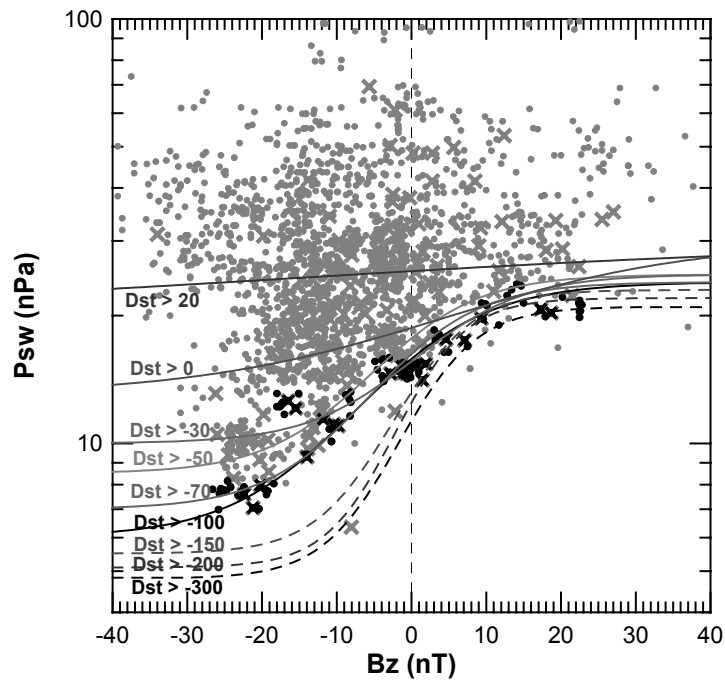

Figure 11. Scatter plot of total solar wind pressure *Psw* versus IMF *Bz* in aGSM for the GMCs (gray crosses) and MSh intervals (gray circles) collected in the range of *Dst* > -100 nT. The best fit of envelope boundary is indicated by black solid curve. The GMCs and MSh intervals selected for the boundary fitting are indicated by black symbols. For comparison, others envelope boundaries derived in various *Dst* ranges are presented by different lines. One can see a fast growing of the asymptotic pressure $P_{min}$ with *Dst*. At large positive *Dst* the $P_{min}$ is approaching to $P_{max}$.

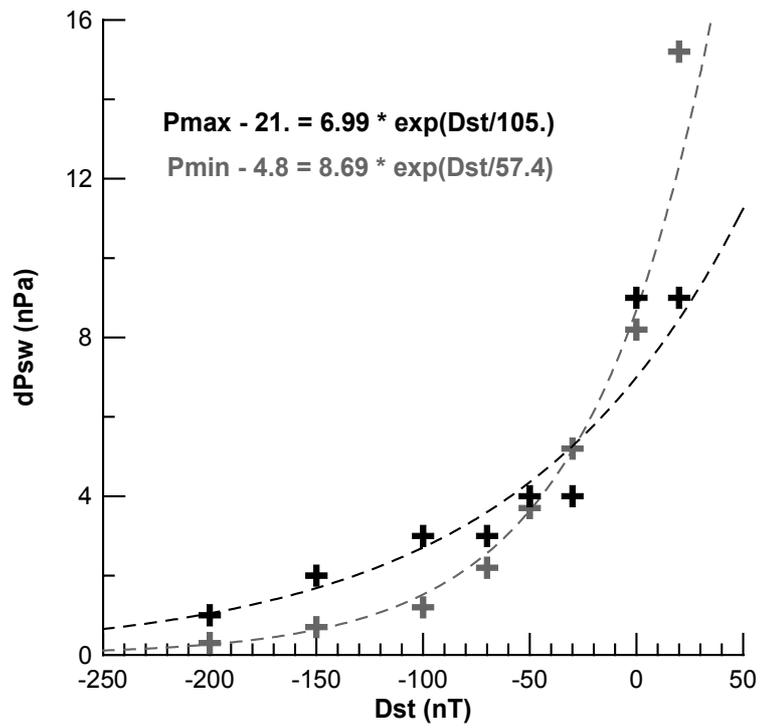

Figure 12. Maximal $P_{max}$ and minimal $P_{min}$ asymptotic pressures of the envelope boundary obtained in noon sector for various ranges of *Dst*. The dashed lines correspond to best fit of the pressures by an exponential function of *Dst*.

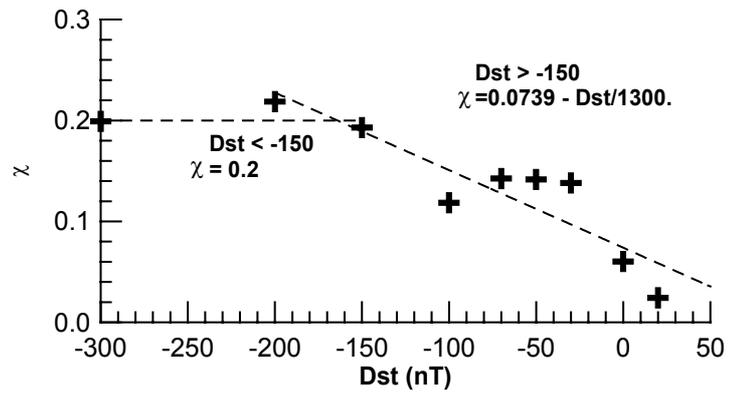

Figure 13. The steepness of the envelope boundary $\chi$ calculated in various ranges of *Dst*. The dashed lines indicate two linear approximations for $-300 < Dst < -150$ nT and for $Dst > -150$ nT.